\documentclass{PoS}

\usepackage{ulem}
\usepackage{color}

\makeatletter
\def\simleq{\mathrel{\mathpalette\gl@align<}}
\def\simgeq{\mathrel{\mathpalette\gl@align>}}
\def\gl@align#1#2{\lower.6ex\vbox{\baselineskip\z@skip\lineskip\z@
     \ialign{$\m@th#1\hfill##\hfil$\crcr#2\crcr\sim\crcr}}}
\makeatother

\title{%
Baryon interactions from lattice QCD with physical masses
-- Overview and $S = 0, -4$ sectors --
}

\ShortTitle{%
Baryon interactions from LQCD with physical masses
-- Overview and $S = 0, -4$ sectors --
}

\author{
\speaker{Takumi~Doi},$^a$ 
Sinya~Aoki,$^{abc}$
Shinya~Gongyo,$^{ad}$
Tetsuo~Hatsuda,$^{ae}$
Yoichi~Ikeda,$^{af}$
Takashi~Inoue,$^{ag}$
Takumi~Iritani,$^{a}$
Noriyoshi~Ishii,$^{af}$
Takaya~Miyamoto,$^{ab}$
Keiko~Murano,$^{af}$
Hidekatsu~Nemura,$^{ac}$
and
Kenji~Sasaki$^{ab}$ \\
\llap{$^a$} Theoretical Research Division, Nishina Center, RIKEN, Wako 351-0198, Japan\\
\llap{$^b$} Yukawa Institute for Theoretical Physics, Kyoto University, Kyoto 606-8502, Japan\\
\llap{$^c$} Center for Computational Sciences, University of Tsukuba, Ibaraki 305-8571, Japan\\
\llap{$^d$} CNRS, Laboratoire de Math\'ematiques et Physique Th\'eorique, Universit\'ede Tours, 37200 France\\
\llap{$^e$} iTHEMS Program and iTHES Research Group, RIKEN, Wako 351-0198, Japan\\
\llap{$^f$} Research Center for Nuclear Physics (RCNP), Osaka University, Osaka 567-0047, Japan\\
\llap{$^g$} Nihon University, College of Bioresource Sciences, Kanagawa 252-0880, Japan\\
E-mail: \email{doi@ribf.riken.jp}
}

\abstract{%
Nuclear forces and hyperon forces are studied by lattice QCD.
Simulations are performed with (almost) physical quark masses,
$m_\pi \simeq 146$ MeV and $m_K \simeq 525$ MeV,
where $N_f=2+1$ nonperturbatively ${\cal O}(a)$-improved Wilson quark action with stout smearing
and Iwasaki gauge action are employed on the lattice of
$(96a)^4 \simeq (8.1\mbox{fm})^4$ with $a^{-1} \simeq 2.3$ GeV.
In this report, we give the overview of the theoretical framework and
present the numerical results for two-nucleon forces ($S=0$) and two-$\Xi$ forces ($S=-4$).
Central forces are studied in $^1S_0$ channel,
and central and tensor forces are obtained in $^3S_1$-$^3D_1$ coupled channel analysis.
}

\FullConference{34th annual International Symposium on Lattice Field Theory\\
		24-30 July 2016\\
		University of Southampton, UK}

\begin{document}

\vspace*{-12mm}
\section{Introduction}
\vspace*{-1mm}
\label{sec:intro}

One of the most challenging issues in particle and nuclear physics is 
to unravel the origin of 
baryon forces
based on the fundamental theory, quantum chromodynamics (QCD).
The precise information on nuclear and hyperon forces 
serve as the key ingredients to calculate properties of nuclei and dense matter
and the structure of neutron stars.
While so-called realistic nuclear forces
have been obtained using experimental scattering phase shifts,
their connection to QCD is yet to be established.
Hyperon forces suffer from large uncertainties
since scattering experiments with hyperon(s) are very difficult
due to the short life time of hyperons.
Under these circumstances, it is most desirable 
to carry out the first-principles calculations of baryon forces
by lattice QCD.

We study baryon forces using a novel theoretical framework, 
HAL QCD method,
in which the interaction kernels (so-called ``potentials'') 
are determined from Nambu-Bethe-Salpeter (NBS) correlators 
on a lattice~\cite{Ishii:2006ec, HALQCD:2012aa, Aoki:2012tk}.
(For the application of obtained lattice baryon forces to
 the equation of state of dense matter, the structure of neutron stars
 and the properties of nuclei, see Refs.~\cite{Inoue:2013nfe, McIlroy:2017ssf}.)
The significant advantage of HAL QCD method over the traditional approach
(so-called ``direct'' calculations~\cite{Yamazaki:2015asa, Orginos:2015aya, Berkowitz:2015eaa})
is that the baryon-baryon interactions can be extracted
without relying on the ground state saturation~\cite{HALQCD:2012aa}.
In fact, since typical excitation energy in multi-baryon systems 
is one to two orders of magnitude smaller than ${\cal O}(\Lambda_{\rm QCD})$ due to
the existence of elastic excited states,
the results from the ``direct'' method~\cite{Yamazaki:2015asa, Orginos:2015aya, Berkowitz:2015eaa},
which rely on the ground state saturation,
become generally unreliable~\cite{Iritani:2016jie, Iritani:2016xmx}.


%

In this paper, we report the latest lattice QCD results for the 
baryon forces obtained at (almost) physical quark masses,
updating our previous results~\cite{Doi:2015oha}.
We first give a brief overview of the theoretical framework
and then present numerical results for 
two-$\Xi$ ($\Xi\Xi$) forces ($S=-4$) and two-nucleon ($NN$) forces ($S=0$)
in parity-even channel.
Central forces are extracted in $^1S_0$ channel,
and central and tensor forces are obtained in $^3S_1$-$^3D_1$ coupled channel analysis.
The results for other baryon forces in the same lattice setup are presented
in Refs.~\cite{Ishii:lat2016}.

\vspace*{-1mm}
\section{Formalism}
\vspace*{-1mm}
\label{sec:formalism}

The key quantity in the HAL QCD method is 
the equal-time NBS wave function.
In the case of a $NN$ system, for instance, it is defined by 
%
$
\phi_W^{NN}(\vec{r}) \equiv 
1/Z_N \cdot
\langle 0 | N(\vec{r},0) N(\vec{0},0) | NN, W \rangle_{\rm in} ,
$
%
where 
$N$ is the nucleon operator
with its wave-function renormalization factor $\sqrt{Z_N}$ 
and
$|NN, W \rangle_{\rm in}$ denotes the asymptotic in-state of the $NN$ system 
at the total energy of $W = 2\sqrt{k^2+m_N^2}$
with 
the asymptotic momentum $k$,
and we consider the elastic region,  $W < W_{\rm th} = 2m_N + m_\pi$.
The most important property of the NBS wave function is that
the asymptotic behavior at $r \equiv |\vec{r}| \rightarrow \infty$ 
is given by
%
$
\phi_W^{NN} (\vec{r}) \propto 
\sin(kr-l\pi/2 + \delta_W^l) / (kr),
$
%
where 
$\delta_W^l$ is the scattering phase shift
with the orbital angular momentum $l$.
Exploiting this feature,
one can define (non-local) $NN$ potential, $U^{NN}(\vec{r},\vec{r}')$,
which is faithful to the phase shifts
through the Schr\"odinger equation~\cite{Ishii:2006ec, HALQCD:2012aa, Aoki:2012tk},
%
%
$
(E_W^{NN} - H_0) \phi_W^{NN}(\vec{r})
= 
\int d\vec{r'} U^{NN}(\vec{r},\vec{r'}) \phi_W^{NN}(\vec{r'}) ,
$
%
where 
$H_0 = -\nabla^2/(2\mu)$ and
$E_W^{NN} = k^2/(2\mu)$ with the reduced mass $\mu = m_N/2$.
It has been also proven that 
$U^{NN}(\vec{r},\vec{r'})$ can be constructed
as to be energy-independent~\cite{Ishii:2006ec,Aoki:2012tk}.

Generally speaking, the NBS wave function 
can be extracted from the 
four-point correlator,
$
G^{NN} (\vec{r},t)
\equiv
\sum_{\vec{R}}
\langle 0 |
          (N(\vec{R}+\vec{r}) N (\vec{R}))(t)\
\overline{(N N)}(t=0)
| 0 \rangle ,
$ 
by isolating the contribution from each energy eigenstate
(most typically by the ground state saturation with $t \rightarrow \infty$).
%
%
Such a procedure, however, is practically almost impossible,
due to the existence of nearby elastic scattering states.
In fact, the typical excitation energy is
as small as ${\cal O}(1)-{\cal O}(10)$ MeV,
which is estimated by the empirical binding energies and/or
the discretization in spectrum by the finite volume, $\sim (2\pi/L)^2 / m_N$.
Correspondingly, ground state saturation 
requires 
$t \simgeq {\cal O}(10)-{\cal O}(100)$ fm, 
which is far beyond reach considering that
signal/noise is exponentially suppressed in terms of $t$.
%

\begin{figure}[t]
\begin{minipage}{0.48\textwidth}
\begin{center}
\vspace*{-4mm}
\includegraphics[angle=0,width=0.85\textwidth]{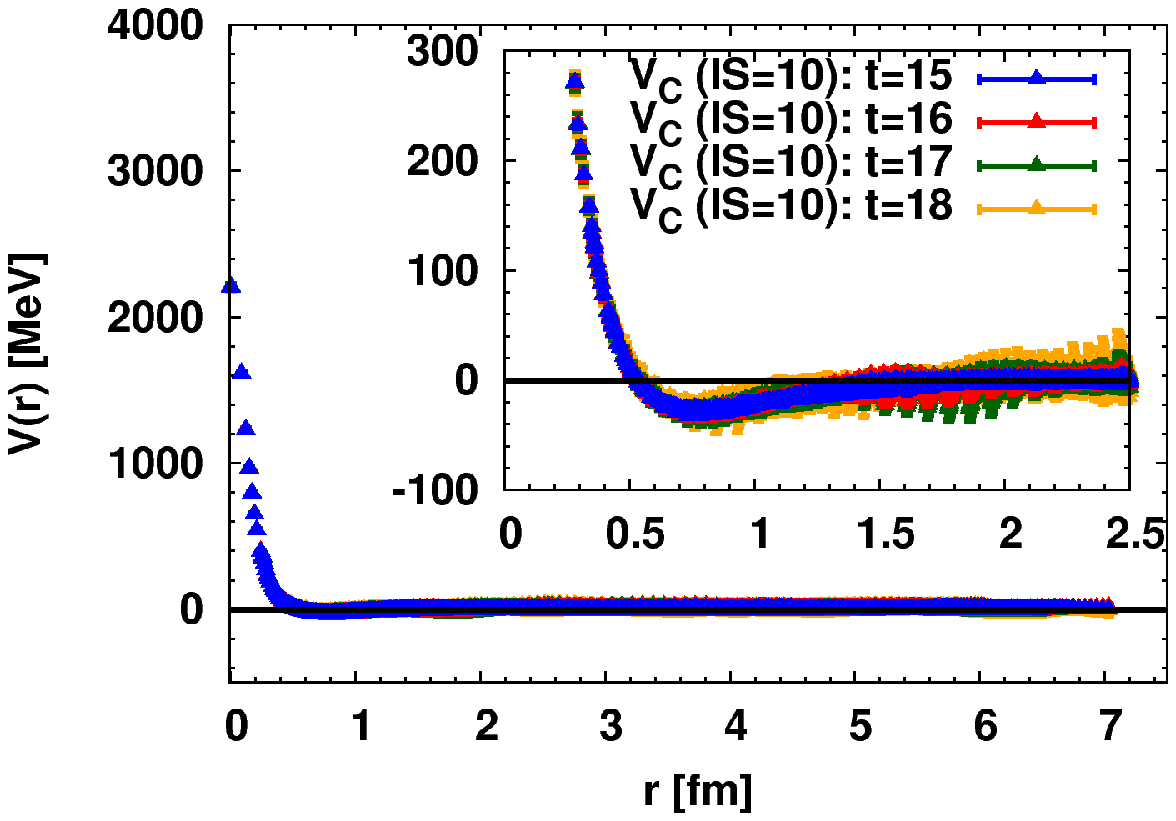}
\vspace*{-2mm}
\caption{
\label{fig:pot:XiXi:1S0:cen}
$\Xi\Xi$ central force $V_C(r)$ in $^1S_0$ $(I=1)$ channel
obtained at $t = 15-18$.
}
\end{center}
\end{minipage}
\hfill
\begin{minipage}{0.48\textwidth}
\begin{center}
\vspace*{-4mm}
\includegraphics[angle=0,width=0.85\textwidth]{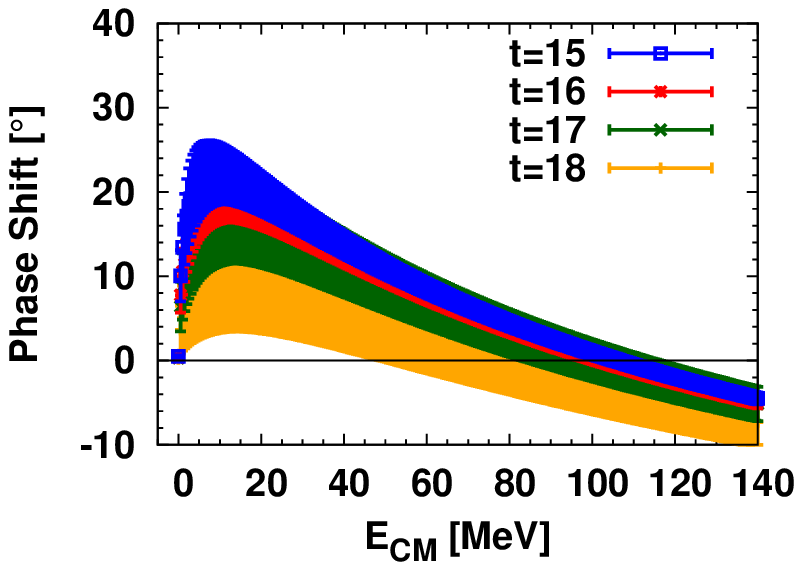}
\vspace*{-2mm}
\caption{
\label{fig:phase:XiXi:1S0:cen}
Phase shifts in $\Xi\Xi (^1S_0)$ $(I=1)$ channel
obtained at $t = 15-18$.
}
\end{center}
\end{minipage}
\end{figure}

Recently, the breakthrough on this issue was achieved 
by the time-dependent HAL QCD method~\cite{HALQCD:2012aa}.
The crucial point is that,
since $U^{NN}(\vec{r},\vec{r'})$ is energy-independent, 
one can extract the signal thereof 
even from elastic excited states.
More specifically, 
the following ``time-dependent'' Schr\"odinger equation holds
even without the ground state saturation,
\begin{eqnarray}
\left( 
- \frac{\partial}{\partial t} 
+ \frac{1}{4m_N} \frac{\partial^2}{\partial t^2} 
- H_0
\right)
R^{NN}(\vec{r},t) 
=
\int d\vec{r'} U^{NN}(\vec{r},\vec{r'}) R^{NN}(\vec{r'},t) ,
\label{eq:Sch_2N:tdep}
\end{eqnarray}
%
where
$R^{NN}(\vec{r},t) \equiv G^{NN} (\vec{r},t) e^{2m_Nt}$.
While it is still necessary to suppress the contaminations from inelastic states,
it can be fulfilled by much easier condition,
$t \simgeq (W_{\rm th} - W)^{-1} \sim {\cal O}(1)$ fm.
This is in contrast to the direct calculations,
which inevitably rely on the ground state saturation.
Note that while ``plateau-like'' structures in the effective energy shifts often appear at $t\sim {\cal O}(1)$ fm
and are customarily used in the previous direct calculations~\cite{Yamazaki:2015asa, Orginos:2015aya, Berkowitz:2015eaa},
they cannot be distinguished from the fake plateaux (called ``mirage'' in Ref.~\cite{Iritani:2016jie})
and thus are generally unreliable~\cite{Iritani:2016jie, Iritani:2016xmx}.

The time-dependent HAL QCD method is also essential to study coupled channel systems.
Coupled channel effects play an important role in hyperon forces,
e.g., in the $\Lambda N$-$\Sigma N$ system and $\Lambda \Lambda$-$N \Xi$-$\Sigma \Sigma$ system.
The master formula 
for a coupled channel system is given by~\cite{Aoki:2011gt}
%
\begin{eqnarray}
\biggl( -\frac{\partial}{\partial t} - H^{\alpha}_{0} \biggr)
R^{\alpha \beta}(\vec{r}, t) 
=  
\sum_{\gamma} \Delta^{\alpha \gamma}
\int d\vec{r'}  U^{\alpha \gamma}(\vec{r},\vec{r'}) 
R^{\gamma \beta}(\vec{r'}, t) ,
\label{t-dep_local}
\end{eqnarray}
where $\alpha, \beta, \gamma$ denote labels for channels,
$R^{\alpha\beta}$ normalized four-point correlator in $\alpha$ ($\beta$) channel at the sink (source),
$H^{\alpha}_{0}=-\nabla^{2}/2\mu^{\alpha}$ with the reduced mass $\mu^{\alpha}=m^{\alpha}_{1} m^{\alpha}_{2} /(m^{\alpha}_{1} + m^{\alpha}_{2})$,
$\Delta^{\alpha \gamma}=e^{(m_{1}^{\alpha}+m_{2}^{\alpha}) t}/e^{(m_{1}^{\gamma}+m_{2}^{\gamma}) t}$,
and subscripts $1, 2$ labels for (two) hadrons in the channel.
For simplicity, 
relativistic correction terms
are omitted. 
The advantage of the coupled channel approach in HAL QCD method 
is also shown in the study of the tetraquark candidate, $Z_c(3900)$~\cite{Ikeda:2016zwx}.

The computational challenge in lattice QCD for multi-baryon systems 
is that enormous computational resources are required for the calculation of correlators.
The reasons are that 
(i) the number of 
Wick contractions
grows  
factorially with mass number $A$, and
(ii) the number of 
color/spinor contractions grows exponentially for larger $A$.
On this point, we recently develop a novel algorithm
called the unified contraction algorithm (UCA),
in which two contractions (i) and (ii) are unified in a systematic way~\cite{Doi:2012xd}.
This algorithm significantly reduces the computational cost
and play a crucial role in our simulations.

\vspace*{-1mm}
\section{Lattice QCD setup}
\vspace*{-1mm}
\label{sec:setup}

$N_f = 2+1$ gauge configurations are generated on the $96^4$ lattice
with the Iwasaki gauge action at $\beta = 1.82$ and 
nonperturbatively ${\cal O}(a)$-improved Wilson quark action with $c_{sw} = 1.11$ 
and APE stout smearing with $\alpha = 0.1$, $n_{\rm stout} = 6$.
About 2000 trajectories are generated after the thermalization,
and preliminary studies show that $a^{-1} \simeq 2.333$ GeV ($a \simeq 0.0846$ fm)
and $m_\pi \simeq 146$ MeV, $m_K \simeq 525$ MeV. 
The lattice size, $La \simeq 8.1$ fm, is sufficiently large
to accommodate two baryons on a box.
For further details on the gauge configuration generation,
see Ref.~\cite{Ishikawa:2015rho}.

The measurements of NBS correlators are performed at the unitary point,
where the block solver~\cite{Boku:2012zi} is used for the quark propagator
and unified contraction algorithm~\cite{Doi:2012xd} is used for the contraction.
The computation for the measurements (including I/O)
achieves $\sim$ 25\% efficiency, or $\sim$ 65 TFlops sustained on 2048 nodes of K computer.
For two-octet baryon forces, we calculate all 52 channels relevant in parity-even channel.
We employ wall quark source with Coulomb gauge fixing,
where
the periodic (Dirichlet) boundary condition is used for spacial (temporal) directions
and forward and backward propagations are averaged to reduce the statistical fluctuations.
We pick 1 configuration per each 5 trajectories,
and we make use of the rotation symmetry to increase the statistics.
The total statistics in this report amounts to
414 configurations $\times$ 4 rotations $\times$ 48 wall sources,
binned by 46 configurations.
%
%
Baryon forces are determined in  $^1S_0$ and $^3S_1$-$^3D_1$ channels.
We perform the velocity expansion~\cite{Aoki:2012tk} in terms of 
the non-locality of potentials,
and obtain the leading order potentials, i.e., central and tensor forces.
In this preliminary analysis shown below, 
the term which corresponds to the relativistic effects
($\partial^2 / \partial t^2$-term in Eq.~(\ref{eq:Sch_2N:tdep}))
is neglected.

\vspace*{-1mm}
\section{$\Xi\Xi$ systems ($S=-4$ channel)}
\vspace*{-1mm}
\label{sec:XiXi}

\begin{figure}[t]
\begin{minipage}{0.48\textwidth}
\begin{center}
\vspace*{-4mm}
\includegraphics[angle=0,width=0.85\textwidth]{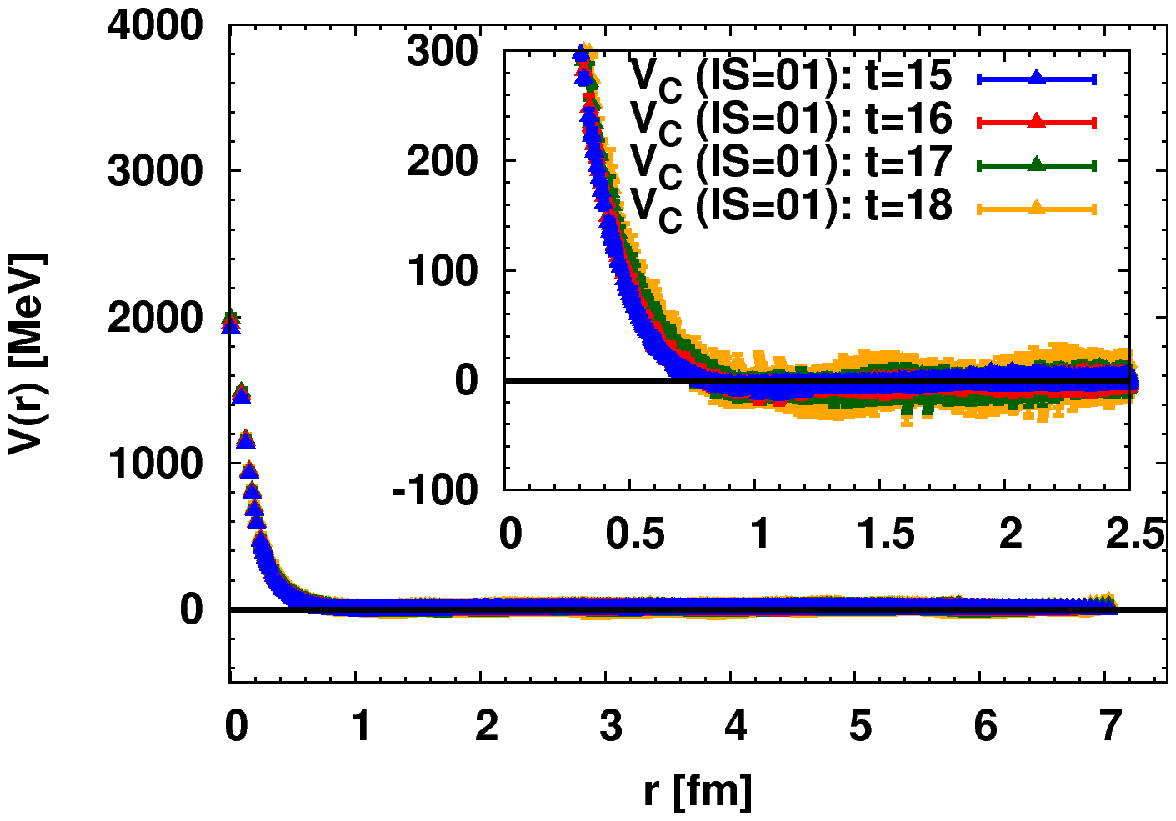}
\vspace*{-2mm}
\caption{
\label{fig:pot:XiXi:3S1:cen}
$\Xi\Xi$ central force $V_C(r)$ in $^3S_1$-$^3D_1$ $(I=0)$ channel
obtained at $t = 15-18$.
}
\end{center}
\end{minipage}
\hfill
\begin{minipage}{0.48\textwidth}
\begin{center}
\vspace*{-4mm}
\includegraphics[angle=0,width=0.85\textwidth]{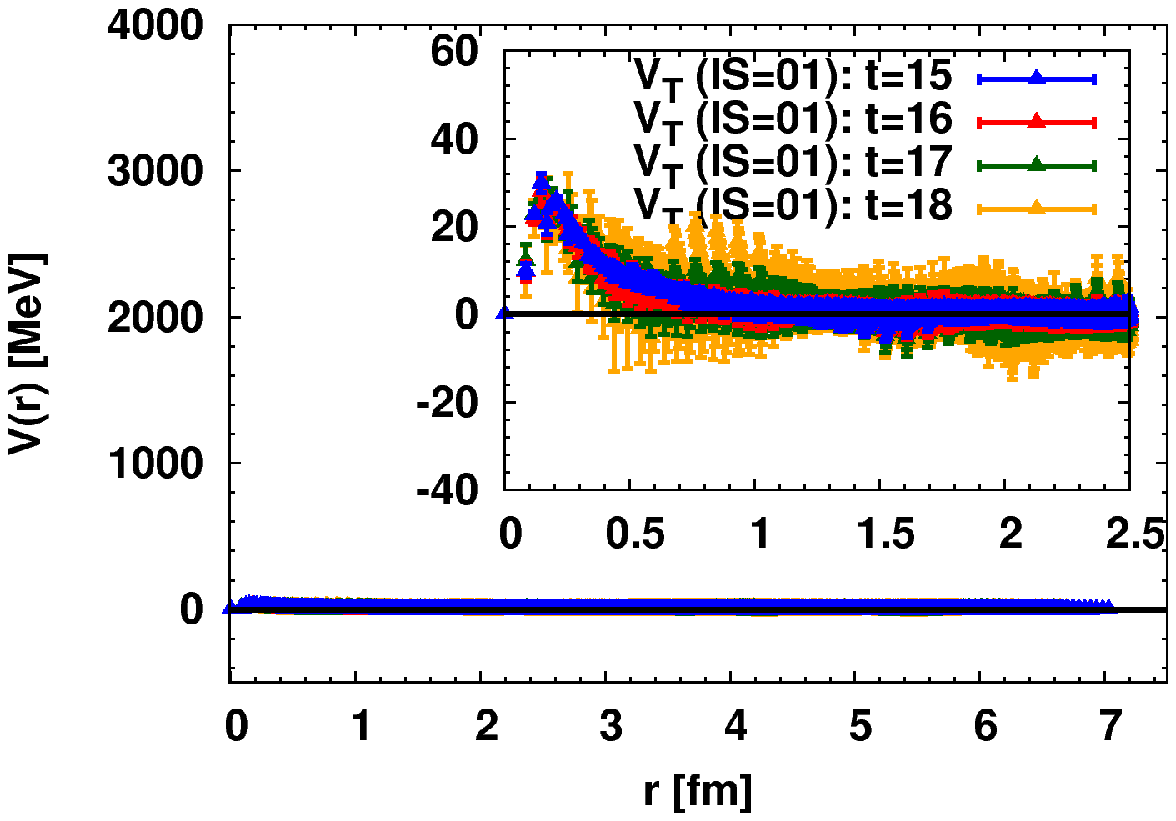}
\vspace*{-2mm}
\caption{
\label{fig:pot:XiXi:3S1:ten}
$\Xi\Xi$ tensor force $V_T(r)$ in $^3S_1$-$^3D_1$ $(I=0)$ channel
obtained at $t = 15-18$.
}
\end{center}
\end{minipage}
\end{figure}

Let us first consider the $\Xi\Xi$ system in $^1S_0$ (iso-triplet) channel.
This channel belongs to the 27-plet 
in flavor SU(3) classification
as does the $NN (^1S_0)$ system.
Therefore, 
the $\Xi\Xi (^1S_0)$ interaction serves as a good ``doorway'' to probe 
the $NN (^1S_0)$ interaction.
In addition,
since the strong attraction in $NN (^1S_0)$ makes a ``dineutron'' nearly bound,
it has been attracting interest whether 
the 27-plet interaction with the SU(3) breaking effects
forms a bound $\Xi\Xi (^1S_0)$ state or not~\cite{Haidenbauer:2014rna}.

In Fig.~\ref{fig:pot:XiXi:1S0:cen},
we show the lattice QCD results
for the central force $V_C(r)$
in the $\Xi\Xi (^1S_0)$ channel.
We observe a clear signal of 
the mid- and long-range attraction as well as the repulsive core at short-range,
resembling the phenomenological potential in $NN(^1S_0)$ system.
Within statistical fluctuations,
the results are found to be consistent with each other in the range $t = 15-18$,
which suggests that the contaminations from inelastic excited states are suppressed
and higher-order terms in the velocity expansion are small.
%
%
Shown in Fig.~\ref{fig:phase:XiXi:1S0:cen}
are the 
corresponding
phase shifts in terms of the center-of-mass energy.
The results indicate that the interaction is strongly attractive at low energies 
while it is not sufficient to form a bound $\Xi\Xi (^1S_0)$ state.
It is desirable to examine this observation in experiments by, e.g., heavy-ion collisions.

We next consider the $\Xi\Xi$ system in $^3S_1$-$^3D_1$ (iso-singlet) channel.
This channel belongs to the 10-plet in flavor SU(3),
a unique representation with hyperon degrees of freedom.
By solving the coupled channel Schr\"odinger equation with 
NBS correlators, we determine the central and tensor forces.
In Figs.~\ref{fig:pot:XiXi:3S1:cen} and ~\ref{fig:pot:XiXi:3S1:ten},
we show the central and tensor forces, respectively.
For the central force, 
we observe the strong repulsive core,
which 
can be understood from the viewpoint of
the quark Pauli blocking effect~\cite{Aoki:2012tk, Oka:1986fr}.
There also exists an indication of a weak attraction at mid range which
may reflect the effect of small attractive one-pion exchange potential (OPEP).
We observe that the $\Xi\Xi$ tensor force (Fig.~\ref{fig:pot:XiXi:3S1:ten})
has opposite sign and is weaker 
compared to 
the $NN$ tensor forces (Fig.~\ref{fig:pot:NN:3S1:ten}).
This could be understood by the phenomenological 
one-boson exchange potentials,
where $\eta$ gives weaker and positive tensor forces
and $\pi$ gives much weaker and negative tensor forces
with flavor SU(3) meson-baryon couplings
together with $F/D$ ratio by SU(6) quark model.
Further studies with larger statistics are currently underway.

\vspace*{-1mm}
\section{$NN$ systems ($S=0$ channel)}
\vspace*{-1mm}
\label{sec:NN}

\begin{figure}[t]
\begin{minipage}{0.48\textwidth}
\begin{center}
\vspace*{-4mm}
\includegraphics[angle=0,width=0.85\textwidth]{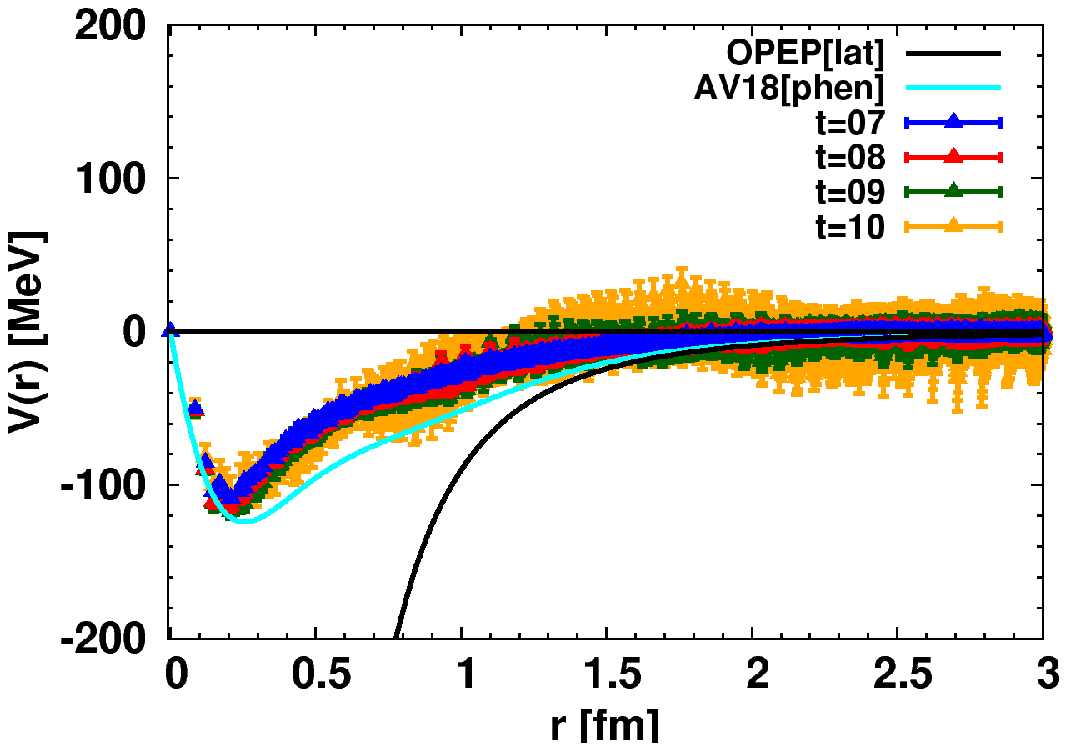}
\vspace*{-2mm}
\caption{
\label{fig:pot:NN:3S1:ten}
$NN$ tensor force $V_T(r)$ in $^3S_1$-$^3D_1$ $(I=0)$ channel
obtained at $t = 7-10$,
together with bare OPEP 
and AV18 phenomenological potential.
}
\end{center}
\end{minipage}
\hfill
\begin{minipage}{0.48\textwidth}
\begin{center}
\vspace*{-9mm}
\includegraphics[angle=0,width=0.85\textwidth]{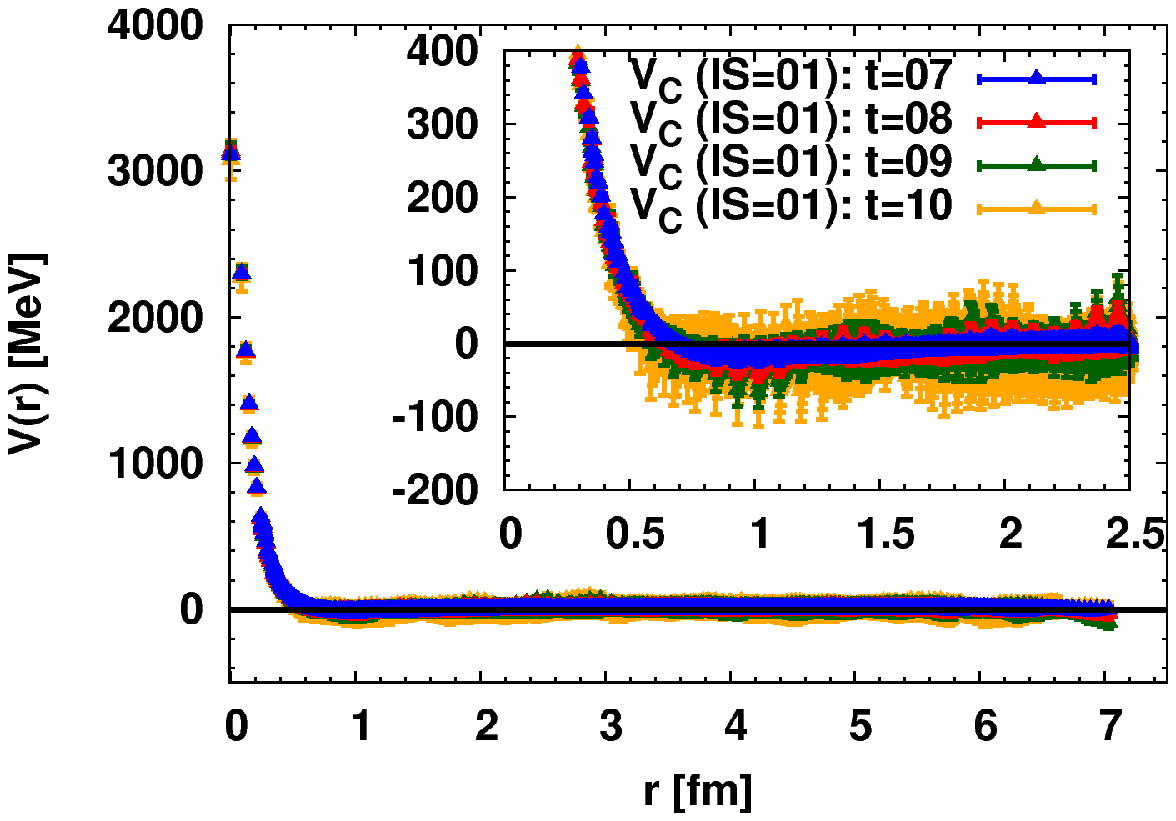}
\vspace*{-2mm}
\caption{
\label{fig:pot:NN:3S1:cen}
$NN$ central force $V_C(r)$ in $^3S_1$-$^3D_1$ $(I=0)$ channel
obtained at $t = 7-10$.
}
\end{center}
\end{minipage}
\end{figure}

Let us begin with the $^3S_1$-$^3D_1$ (iso-singlet) channel.
In Fig.~\ref{fig:pot:NN:3S1:ten}, we show the tensor force $V_T(r)$ obtained at $t = 7-10$.
A strong tensor force with the long-range tail is clearly visible.
Compared to the lattice tensor forces obtained with heavier quark masses~\cite{Aoki:2012tk},
the range of interaction is found to be longer.
We also make a qualitative comparison with bare OPEP and the phenomenological tensor force (AV18)
in Fig.~\ref{fig:pot:NN:3S1:ten},
while a quantitative comparison requires a study in terms of phase shifts
which will be presented elsewhere.
The tail structures are found to be rather similar among these three potentials.
Overall behaviors including the suppression at short-range 
compare to bare OPEP are also similar between lattice potentials and AV18.
Since it is the tensor force which plays the most crucial role in 
the binding of deuteron,
this is a very 
encouraging 
result.
In order to suppress contaminations from inelastic states,
it is desirable to take larger $t$ by increasing the statistics, which is in progress.
In Fig.~\ref{fig:pot:NN:3S1:cen}, we show the central force $V_C(r)$ in $^3S_1$-$^3D_1$ channel.
While the results suffer from much larger statistical fluctuations,
the repulsive core at short-range is clearly observed
and mid- and long-range attraction is obtained as well.

\begin{figure}[t]
\begin{minipage}{0.48\textwidth}
\begin{center}
\vspace*{-4mm}
\includegraphics[angle=0,width=0.85\textwidth]{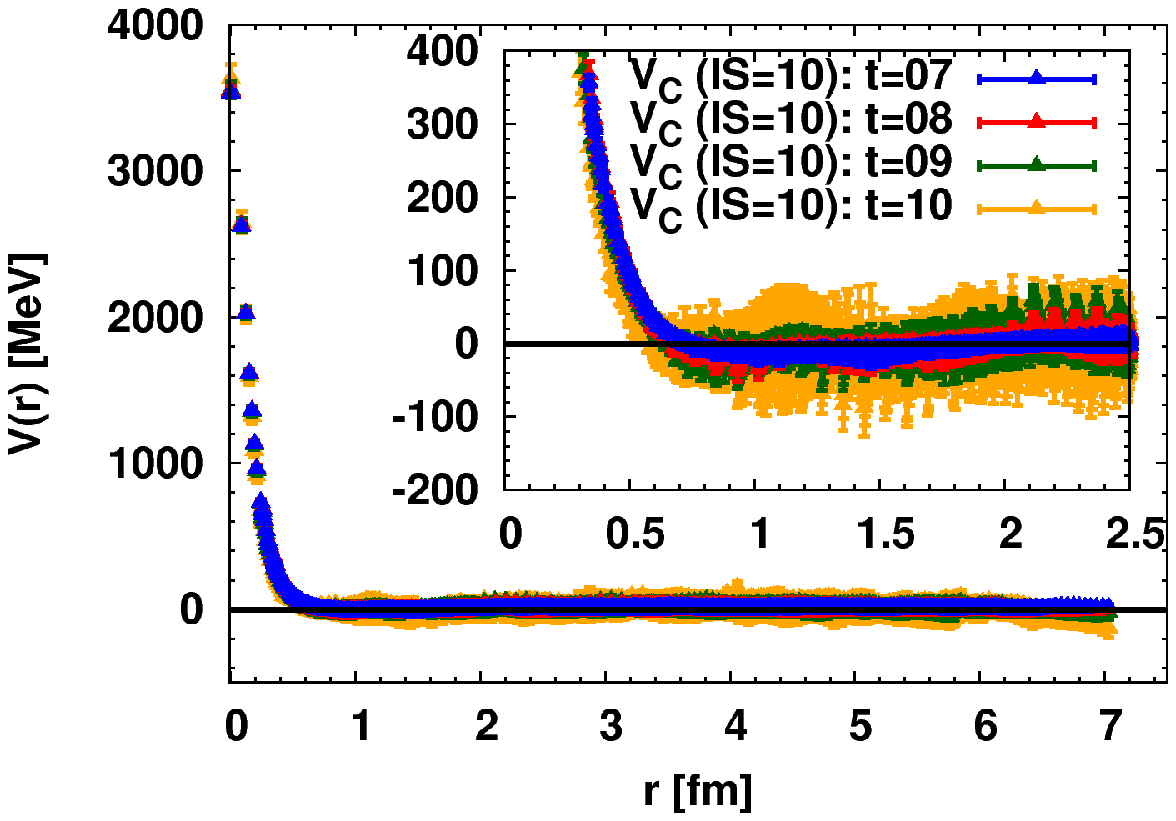}
\vspace*{-2mm}
\caption{
\label{fig:pot:NN:1S0:cen}
$NN$ central force $V_C(r)$ in $^1S_0$ $(I=1)$ channel
obtained at $t = 7-10$.
}
\end{center}
\end{minipage}
\hfill
\begin{minipage}{0.48\textwidth}
\begin{center}
\vspace*{-4mm}
\includegraphics[angle=0,width=0.85\textwidth]{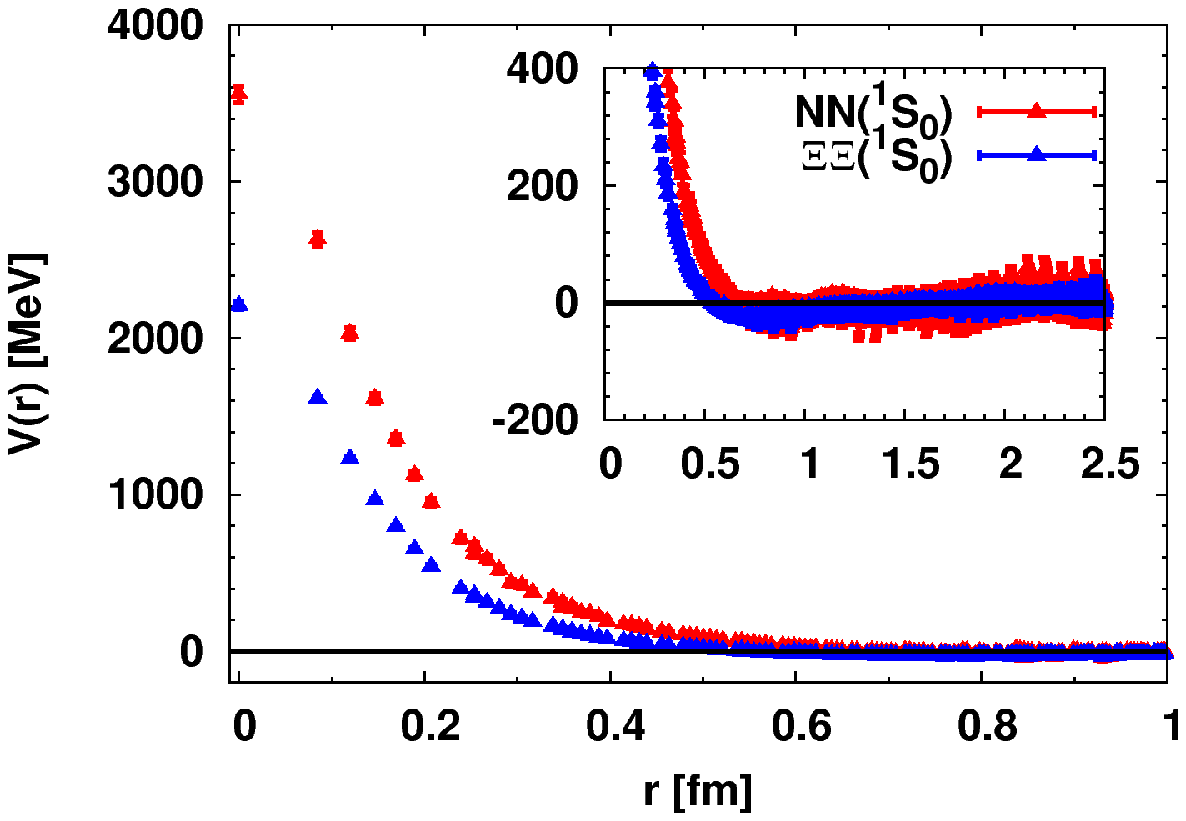}
\vspace*{-2mm}
\caption{
\label{fig:pot:NNvsXiXi:1S0:cen}
Comparison between 
central forces in $NN(^1S_0)$ at $t=9$ (red) and $\Xi\Xi(^1S_0)$ at $t=18$ (blue).
}
\end{center}
\end{minipage}
\end{figure}

We then consider the $^1S_0$ (iso-triplet) channel.
Shown in Fig.~\ref{fig:pot:NN:1S0:cen} is the obtained central force $V_C(r)$ for $NN(^1S_0)$.
As is the case for the central force in $^3S_1$-$^3D_1$ channel,
the results suffer from large statistical fluctuations.
Yet, the repulsive core at short-range is observed
and mid- and long-range attraction is obtained as well.
In Fig.~\ref{fig:pot:NNvsXiXi:1S0:cen}, we compare 
$NN(^1S_0)$ at $t=9$ (red) and $\Xi\Xi(^1S_0)$ at $t=18$ (blue).
As noted before, both channels 
belong to the 27-plet in flavor SU(3) classification,
and the difference dictates the SU(3) breaking effect.
We observe that the repulsive core in $NN(^1S_0)$ is more enhanced 
than $\Xi\Xi(^1S_0)$,
which 
can be understood from the one-gluon exchange picture.

\vspace*{-1mm}
\section{Summary}
\vspace*{-1mm}
\label{sec:summary}

We have reported the lattice QCD studies
for baryon interactions 
with (almost) physical quark masses,
$m_\pi \simeq 146$ MeV and $m_K \simeq 525$ MeV
on a large lattice box $(96 a)^4 \simeq (8.1 {\rm fm})^4$.
Baryon forces have been calculated from 
NBS 
correlators
in the (time-dependent) HAL QCD method.


Preliminary results for $\Xi\Xi$ and $NN$ interactions have been presented.
In the $\Xi\Xi (^1S_0)$ channel, a strong attraction is obtained,
although it is not strong enough to form a bound state.
In the $\Xi\Xi$ ($^3S_1$-$^3D_1$) channel, 
we have observed the strong repulsive core in the central force.
Tensor force is found to be weak and have an opposite sign compared to the $NN$ tensor force.
For $NN$ forces,
a clear signal for the strong tensor force has been obtained.
Repulsive cores as well as mid- and long-range attractions have been observed 
in central forces in both $^1S_0$ and $^3S_1$-$^3D_1$ channels.
Repulsive core in $NN(^1S_0)$ is found to be stronger than $\Xi\Xi(^1S_0)$.
These observations have interesting physical implications 
from the point of view of quark Pauli blocking effect
and phenomenological models of baryon forces.
Studies in terms of phase shifts with increased statistics are in progress.

\vspace*{-1mm}
\section*{Acknowledgments}
\vspace*{-1mm}

We thank members of PACS Collaboration for the gauge configuration generation.
The lattice QCD calculations have been performed on the K computer at RIKEN, AICS
(hp120281, hp130023, hp140209, hp150223, hp150262, hp160211),
HOKUSAI FX100 computer at RIKEN, Wako (G15023, G16030)
and HA-PACS at University of Tsukuba (14a-20, 15a-30).
We thank ILDG/JLDG~\cite{conf:ildg/jldg}
which serves as an essential infrastructure in this study.
This work is supported in part by 
MEXT Grant-in-Aid for Scientific Research (JP15K17667),
SPIRE (Strategic Program for Innovative REsearch) Field 5 project
and
"Priority Issue on Post-K computer" (Elucidation of the Fundamental Laws and Evolution of the Universe).


\end{document}